\documentclass[12pt,fleqn]{article}
\usepackage{xiiiemc}
\usepackage{natbib}
\usepackage{fancyhdr}
\usepackage{fancyvrb}
\usepackage{color}
\usepackage{wallpaper} 
\usepackage{titlesec}   
\usepackage{float} 	
\usepackage[usenames,dvipsnames]{xcolor}
\usepackage{hyperref} 	
\hypersetup{
	pdfcreator={LaTeX with abnTeX2},
	pdfkeywords={abnt}{latex}{abntex}{USPSC}{trabalho acadêmico}, 
	colorlinks=true,       		
	linkcolor=blue,          	
	citecolor=blue,        		
	filecolor=magenta,      		
	urlcolor=blue,
	allbordercolors=black,
	bookmarksdepth=4
}
\usepackage{tocloft}
\titleformat{\section}
  {\normalfont\bfseries}{\thesection.}{0.5em}{}
 
\renewcommand\thesection{\arabic{section}}

\let\OLDthebibliography\thebibliography
\renewcommand\thebibliography[1]{\OLDthebibliography{#1} \setlength{\parskip}{0pt}\setlength{\itemsep}{0pt plus 0.3ex}}

\usepackage{grffile}
\usepackage{afterpage}

\title{AUDIOVISUAL ANALYTICS VOCABULARY AND ONTOLOGY (AAVO): INITIAL CORE AND EXAMPLE EXPANSION}

\author
    {\rm \begin{tabular}{l} 
    \textbf{Renato Fabbri}$$ - {\textnormal renato.fabbri@gmail.com}\\%
    \textbf{Maria Cristina Ferreira de Oliveira}$$ - {\textnormal cristina@icmc.usp.br}\\
    {\fontsize{11}{0}\selectfont University of São Paulo, Institute of Mathematical and Computer Sciences - São Carlos, SP, Brazil}\vspace*{-0.05cm} \\
  \end{tabular}}

\fancypagestyle{firspagetstyle}
{
	\lhead{}
	\fancyhead[C]{%
		\includegraphics[width=0.9\linewidth]{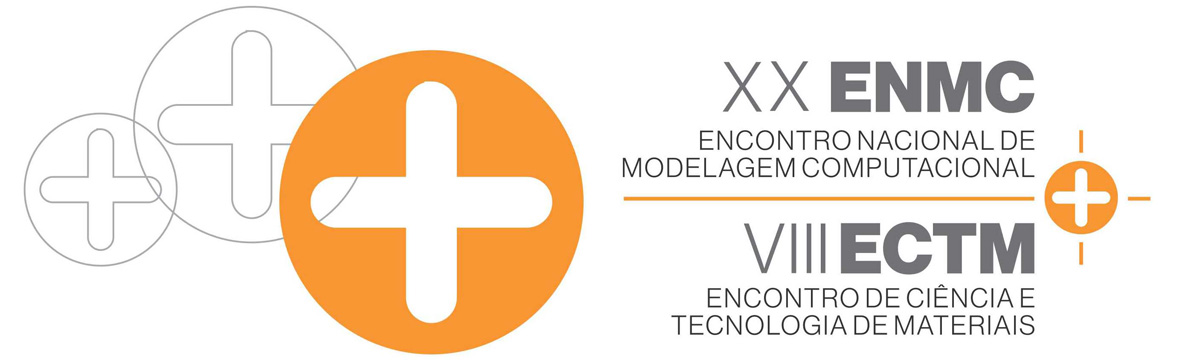}\\%
		{\scriptsize \fontfamily{phv}\fontseries{b}\selectfont \color[rgb]{0.45,0.45,0.45}
		16 a 19 de Outubro de 2017\\
		Instituto Politécnico - Universidade do Estado de Rio de Janeiro\\
		Nova Friburgo - RJ\\
	    }
	}
	\renewcommand{\headrulewidth}{0.0pt}
	\fancyfoot[C]{\footnotesize \parbox{15cm} {\centering  \fontsize{7.5}{0}\selectfont \it Anais do XX ENMC – Encontro Nacional de Modelagem Computacional e VIII ECTM – Encontro de Ciências e Tecnologia de Materiais,  Nova Friburgo, RJ – 16 a 19 Outubro 2017}} 
	\rhead{}
}

\begin{document}
\maketitle

\thispagestyle{firspagetstyle}

\fancyhead[L]{\footnotesize{\fontsize{7.5}{0}\selectfont \it XX ENMC e VIII ECTM\\
	16 a 19 de Outubro de 2017\\
	Instituto Politécnico Universidade do Estado do Rio de Janeiro – Nova Friburgo - RJ\\}}
\renewcommand{\headrulewidth}{0.0pt}
\fancyfoot[C]{\footnotesize \parbox{15cm} {\centering  \fontsize{7.5}{0}\selectfont \it Anais do XX ENMC – Encontro Nacional de Modelagem Computacional e VIII ECTM – Encontro de Ciências e Tecnologia de Materiais,  Nova Friburgo, RJ – 16 a 19 Outubro 2017}} 
\rhead{}

\begin{abstract}
Visual Analytics might be defined as data mining assisted by interactive visual interfaces.
The field has been receiving prominent consideration by researchers, developers and the industry. 
The literature, however, is complex because it involves multiple fields of knowledge
and is considerably recent.
In this article we describe an initial tentative organization of the knowledge in the field as an OWL ontology
and a SKOS vocabulary.
This effort might be useful in many ways that include conceptual considerations and
software implementations.
Within the results and discussions, we expose a core and an example expansion
of the conceptualization, and incorporate design issues that enhance the
expressive power of the abstraction. 
\end{abstract}


\pagestyle{fancy}

\section{INTRODUCTION}\label{sec:intro}
Visual Analytics is often defined as data mining (or the science of analytical reasoning)  facilitated by interactive visual interfaces~\citep{wikiVA,visMaster}.
%
%
%
%
%
From this definition, one can grasp that at least three fields are directly related to visual analytics:
data mining, human-computer interaction, and data visualization.
Each one of these fields are inherently multidisciplinary and known to be considerably complex,
with multiple theories and vast literature. Researchers entering into the field of visual analytics can easily find themselves confused with such a diverse material, possibly presented using  terminology that is not necessarily consistent across the different fields.   

This work reports an initial attempt of organizing the knowledge related to visual analytics
as a SKOS vocabulary and an OWL ontology, i.e. a formalized conceptualization
%
%
%
%
in terms of linked data recommended technologies.
Such a formalization can have multiple uses, including:
\begin{itemize}
	\item Facilitating the introduction of the Visual Analytics subject to non-specialists.
	\item Expressing a concise overview of the field.
	\item Expressing a domain knowledge against which queries might be performed.
%
%
%
%
	\item Relating objects (e.g. data and techniques) within the field or between the field and other domains.
	\item Making inferences about the concepts and objects to which they are related.
\end{itemize}

In the present case, where the formalization is realized as linked data,
the conceptualization allows the queries, inferences, relations, etc.
to be performed also by machines.
Therefore, a software might, for example,  relate a dataset
or analysis methods to specific visualization techniques, in order to assist a user designing a visualization, or
for automated reporting.

Section~\ref{sec:methods} uncovers the methods and technologies
needed for stating and discussing the results in Section~\ref{sec:res}.
Final remarks and future work are in Section~\ref{sec:con}.

\section{METHODS}\label{sec:methods}
The methods described here are standard of the semantic web
to achieve formalized conceptualizations.
Thus, the next sections address the subjects very briefly.
The interested reader should visit the vanilla literature,
especially the W3C recommendations~\citep{w3cld,ldb}.

\subsection{The semantic web}
The semantic web is constituted by data which is linked in the same way
web pages are: through HTTP and URLs.
W3C recommendations provide the main source of protocols and best practices for the field.
The terms `linked data' and `semantic web' are most often used interchangeably.
A distinction might arise in some contexts where one needs to refer to the (linked) data or
the (semantic) web created by all or some portion of linked data, but, generally speaking, the terms are equivalent.
%
%
%
%
%
The main topics of this article are data visualization (or visual analytics)
and semantic web, and all the sections tackle the subject of the semantic web
for it is the framework in which data visualization knowledge is represented.

\subsection{RDF}
The semantic web is built using the Resource Description Framework (RDF).
The RDF data model is based on making statements in the form of triples
(``subject-predicate-object'') and using Unique Resource Identifiers (URIs) for
objects and concepts.
It is also part of the framework to use URIs that are URLs whenever possible,
to enable the data linkage.
Accordingly, one can write:

\begin{Verbatim}[fontsize=\footnotesize]
<http://example.org/people/mary>
	<http://example.org/properties/name> "Mary Shastacian" .
<http://example.org/people/mary>
	<http://example.org/properties/age> "57" .
<http://example.org/people/mary> 
	<http://example.org/properties/likes> 
                <http://example.org/concepts/Reading> .
\end{Verbatim}
\noindent to express that there is a 57 year old person called Mary Shastacian who likes reading.
There are many formats to write/serialize RDF data.
The example above is written in Turtle, which will be the format used throughout this document.

In real settings, when everything is working as recommended,
each of these URIs (that are URLs!) might be accessed 
through HTTP to reach more triples referring to the URI.
From the triples above, one would be able to access triples
describing each of the properties: \texttt{example:properties/name}, \texttt{example:properties/age}, and\\
\texttt{example:properties/likes},
and the concept \texttt{example:concepts/Reading}.
In fact, the triples above would probably be available in the URI/URL:
\texttt{example:people/mary}.
An interesting working example\footnote{Visit \url{http://dbpedia.org/page/Rhesus_macaque}
and click on the concepts and properties to start browsing the web of linked data.} is DBPedia~\citep{dbpedia}.
The process of accessing a URI to find more triples is called \emph{dereferencing the URI} (or simply dereferencing).

\subsection{RDFS}
The Resource Description Framework Schema (RDF Schema or simply RDFS)
is a set of classes and properties for the RDF data model that allows
basic descriptions of ontologies.
It supports taxonomic relations (hypernymy, i.e. relations stating that a concept is more general than another),
bindings of properties to objects and datatypes, and notes (label, comment, see also, is defined by).

\subsection{SKOS}
The Simple Knowledge Organization System (SKOS)
is a data model for representing controlled vocabularies.
SKOS is a W3C recommendation to facilitate publication
and use of vocabularies and is built upon RDF and RDFS.
It is itself a vocabulary for concepts, notation, documentation,
semantic and mapping relations, and collections.

\subsection{OWL}
The Web Ontology Language (OWL) is a language for publishing ontologies on the web.
While RDFS holds basic relations necessary even for very rudimentary organization
knowledge and data, OWL is complex and allows one to formalize elaborate conceptualizations.
Using OWL, an ontology might have properties that are required to satisfy a number or axioms,
and classes that obey restrictions or e.g. are the result of the union of other classes.

\subsection{Interviews with specialists and literature consultation}
The standard approach to design an ontology, according to the literature,
is to interview specialists of the field to which the ontology is related,
or to absorb the established literature, or both.
This work is being developed using both approaches.
The second author is a data visualization specialist who was
interviewed by the first author.
Also, the first author is engaged in acquiring a deeper knowledge of the field.

\section{RESULTS AND DISCUSSION}\label{sec:res}
Using the framework exposed in the previous section,
we elaborated an initial vocabulary and ontology for
visual analytics: the AAVO (Audiovisual Analytics Vocabulary and Ontology).
The inclusion of ``audio'' is both a reminder of the possibilities
available for using audio to represent data and perceive patterns,
and a desirable incorporation of audio to visual analytics given
audiovisual capabilities of current ordinary computers.

The main concepts and their interrelations are presented in
Section~\ref{sec:core} while an example extension is on Section~\ref{sec:ext}.
Section~\ref{sec:voc} holds annotations for the vocabulary which are not promptly
given by the previous sections.
Some of the relations bellow are expressed using very recent 
techniques described in~\cite{enhance}.
Their meaning, though, might be easily inferred.

\subsection{AAVO core}\label{sec:core}
\begin{figure}[!htbp] 
\vspace{-2pt}
\begin{center}
\includegraphics[width=\textwidth]{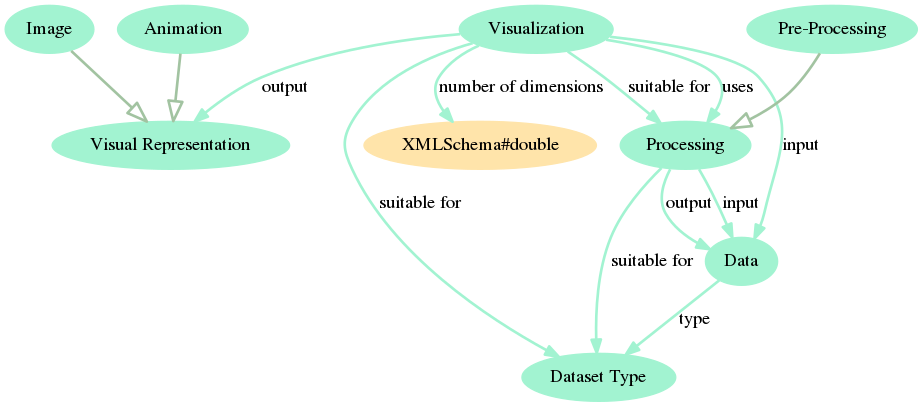}
	\caption{AAVO core, discussed in Section~\ref{sec:core}.
	A Visualization outputs a Visual Representation that can be an Image or an Animation.
	A Visualization is suitable for a Dataset Type and for a Processing routine which transforms
	Data into Data.}
\label{fig:minimum}%
\end{center}
\end{figure}

The core of AAVO is designed to be minimal and hold the following concepts as
depicted in Figure~\ref{fig:minimum}:
\begin{itemize}
	\item Visualization: a technique to generate a Visual Representation from Data.
	\item Visual Representation: a representation of Data by visual cues.
		A Visual Representation can be an Image or an Animation.
	\item Data: a set of values, be them qualitative or quantitative~\citep{wikipData}.
	\item Dataset Type: a type of organization and meaning of data~\citep{munzner}.
	\item Processing: transforms Data into Data. Pre-processing is a kind of Processing.
\end{itemize}

We envision that there should be at least the following concepts in
AAVO core when it reaches maturity:
\begin{itemize}
	\item Hypothesis: a proposed explanation for a phenomenon that might be
		1) given beforehand and amenable to being proved or refuted by an Analysis,
		2) shared by means of an Analysis,
		or 3) presented by means of a Visualization.
	\item Analysis: a set of procedures used to gain understanding about
		Data or a phenomenon.
	\item Task/Purpose/Application: the goal or objective of an Analysis.
\end{itemize}
\noindent These concepts have not yet been included in AAVO core (e.g. Figure~\ref{fig:minimum})
because we are still considering the best way to do so.
From the definitions above,  the question arises:
should we also include Phenomenon among these core concepts?

Other relations that can be added to the core (or to an extension, but
are directly related to the core):
\begin{itemize}
	\item Visualization is a type of Processing.
	\item Visual Representation is a Dataset Type.
	\item Processing ``suitable for'' Data.
	\item Visualization ``number of dimensions'' real (not double as stated for now).
%
%
%
%
%
\end{itemize}

An example of question still left unanswered:
a Visualization only outputs Visual Representation
or can it output other Data(set Type)?
This and many other questions might have a resolution
that are genuinely dependent on the conceptual design of the ontology.


\subsection{AAVO example expansion}\label{sec:ext}
\begin{figure}[!htbp] 
\vspace{-2pt}
\begin{center}
\includegraphics[width=\textwidth]{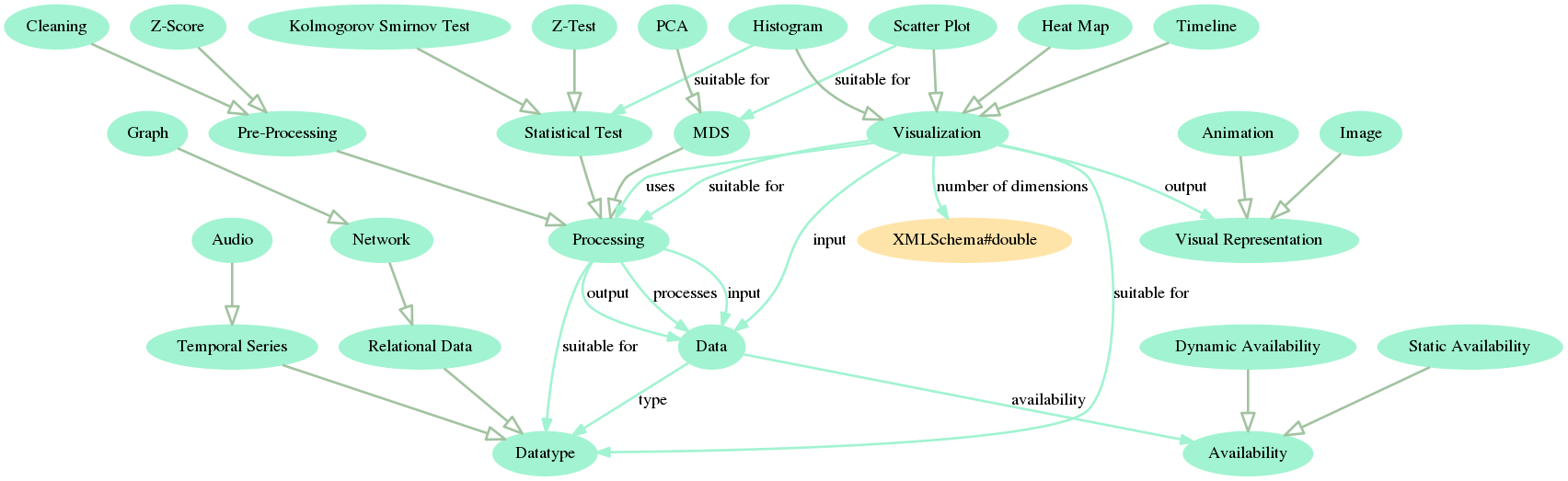}
    \caption{Example incorporation of less general concepts to AAVO: Statistical Test, MDS, Timeline, Z-Score, Network, etc.
	A thorough consideration of this expansion of the AAVO core is at Section~\ref{sec:ext}.
	In case this image is not being properly visualized in paper,
	an online PDF should be at: \url{https://github.com/ttm/aavo/raw/master/latex/article.pdf}.}
\label{fig:exemp}%
\end{center}
\end{figure}

There are many ways in which the AAVO core might be expanded.
Figure~\ref{fig:exemp} is an example expansion.
Concepts were added which are hyponyms to Dataset Type (Temporal Series, Relational Data),
to Pre-Processing (Z-Score, Cleaning), Processing (MDS, Statistical Test)
and Visualization (Heat Map, Histogram, Scatter Plot, Timeline).
Some examples of further subclasses are also added.
A different kind of expansion was achieved by including (Data) Availability
and the less general concepts of Dynamic Availability and Static Availability.
A Graph is regarded as a bare Network without any context or further attributes beyond
nodes and edges.

Ideally, AAVO expansions should reach related fields, such as HCI,
by linking to other existing ontologies (such as DBPedia) or incorporating
enough concepts to then bind and rely in third party conceptualizations.

\subsection{Vocabulary annotations}\label{sec:voc}
Beyond what is made explicit in the previous sections,
there are some aspects of the knowledge and language that
are to be directly added to the SKOS vocabulary.
Examples:
\begin{itemize}
	\item in a dataset, an element is also called: an item, an observation, an individual, a point,
		and even a data point and a data row. 
	\item A graph node is also called: a vertex, and every name that are used to designate an element.
	\item A graph edge is also called: a link, a bond, a line, and a connection.
	\item Z-scores are also called: standard scores, normal scores, standardized variables, and z-values. 
\end{itemize}

\section{CONCLUSIONS AND FURTHER WORK}\label{sec:con}
This initial formalized conceptualization of the AAVO
holds some relations which are not explicitly described by
current literature mainly because of the purposes:
1) of reaching a sound
conceptualization that allows a formalization as linked data;
2) of representing the knowledge in Visual Analytics
to enable inference by machines.
There are other uses for AAVO, uncovered in Section~\ref{sec:intro},
for which conceptual models are available~\citep{munzner,ward}.

Potential further steps include:
\begin{itemize}
	\item the inclusion of the concepts Hypothesis, Analysis, and Task
		into the AAVO core.
	\item Realizing AAVO expansions until the reached concepts can be linked
		to other ontologies that are sound, used and maintained.
	\item Using AAVO to obtain interesting relations by means of automated
		inference and to assist a (audio)visual analytics software.
\end{itemize}

\subsection*{\textit{Acknowledgements}}
The authors thank FAPESP (project 2017/05838-3) for the funding received while researching the topic of this article, the researchers of IFSC/USP and ICMC/USP for the recurrent collaboration in every situation
where we needed directions for investigation.









\begin{thebibliography}{99}
\fontsize{11}{0}\selectfont
\bibitem[Fabbri, 2017]{enhance}
	Fabbri, R. (2017). Enhancements of linked data expressiveness for ontologies.
		Encontro Nacional de Modelagem Computacional 2017 (XX ENMC).
		From \url{https://github.com/ttm/ontologyEnhancements/raw/master/article.pdf}

\bibitem[Heath \& Bizer, 2011]{ldb}
	Heath, T. \& Bizer, C. (2011). Linked Data: Evolving the Web into a Global Data Space (1st edition). Synthesis Lectures on the Semantic Web: Theory and Technology, 1:1, 1-136. Morgan \& Claypool.

\bibitem[Lehmann et al., 2015]{dbpedia}
	Lehmann, J., Isele, R., Jakob, M., Jentzsch, A., Kontokostas, D., Mendes, P. N., ... \& Bizer, C. (2015). DBpedia–a large-scale, multilingual knowledge base extracted from Wikipedia. Semantic Web, 6(2), 167-195.

\bibitem[Munzner, 2014]{munzner}
	Munzner, T. (2014). Visualization analysis and design. CRC press.

\bibitem[W3C, 2010]{w3cld}
	W3C (2017). LINKED DATA CURRENT STATUS, from \url{https://www.w3.org/standards/techs/linkeddata}

\bibitem[Ward et al., 2010]{ward}
	Ward, M. O., Grinstein, G., \& Keim, D. (2010). Interactive data visualization: foundations, techniques, and applications. CRC Press.

\bibitem[Wikipedia, 2017]{wikipData}
	Data. (2017, August 21). In Wikipedia, The Free Encyclopedia. Retrieved
		22:31, August 21, 2017
		, from \url{https://en.wikipedia.org/w/index.php?title=Data&oldid=796493851}
        
\bibitem[Wikipedia, 2017]{wikiVA}
Visual analytics. (2017, July 1). In Wikipedia, The Free Encyclopedia. Retrieved
14:51, August 29, 2017
, from \url{https://en.wikipedia.org/w/index.php?title=Visual_analytics&oldid=788453746}

\bibitem[Ellis \& Mansmann, 2010]{visMaster}
Ellis, G., \& Mansmann, F. (2010). Mastering the information age solving problems with visual analytics. In Eurographics (Vol. 2, p. 5).
\end{thebibliography}
\end{document}